\documentclass[aps,prx,twocolumn,floatfix,showpacs,citeautoscript,longbibliography,hyperlinks]{revtex4-1}
\usepackage{amsmath}
\usepackage{amssymb}
\usepackage{amsfonts}
\usepackage{dsfont}
\usepackage{color}
\usepackage{nicefrac}
\usepackage[T1]{fontenc}
\usepackage[autostyle]{csquotes}
\usepackage[colorlinks=true,citecolor=blue]{hyperref}
\usepackage{graphicx,graphics}

\begin{document}

\title{Designer quantum matter in van der Waals heterostructures}

\author{Jose L. Lado}
\affiliation{Department of Applied Physics, Aalto University, FI-00076 Aalto, Finland.}

\author{Peter Liljeroth}
\affiliation{Department of Applied Physics, Aalto University, FI-00076 Aalto, Finland.}

\begin{abstract}
    Van der Waals materials can be easily combined in lateral and vertical heterostructures, providing an outstanding platform to engineer elusive quantum states of matter.
    However, a critical problem in material science is to establish tangible links between real materials properties and terms that can be cooked up on the model Hamiltonian level to realize different exotic phenomena. Our review aims to do precisely this: we first discuss, in a way accessible to the materials community, what ingredients need to be included in the hybrid quantum materials recipe, and second, we elaborate on the specific materials that would possess the necessary qualities. We will review the well-established procedures for realizing 2D topological superconductors, quantum spin-liquids and flat bands systems, emphasizing the connection between well-known model Hamiltonians and real compounds. We will use the most recent experimental results to illustrate the power of the designer approach.
\end{abstract}

\maketitle 

\section{Introduction}

Two-dimensional materials are at a focus of intense research efforts, with the paradigmatic examples of graphene, hexagonal boron nitride, transition metal dichalcogenides, and transition metal trihalides. The genuine interest in these materials stems from the many high-quality synthesis possibilities, together with the richness of different behaviours. These compounds have been shown to realize properties starting from conventional insulating and metallic behaviour, all the way up to complex many-body ground states such as superconductors and topological insulators.

Besides their intrinsically interesting properties, layered 2D vdW materials can be easily combined in lateral and vertical heterostructures. As the layers only interact via the weak vdW forces, the individual layers can retain their intrinsic properties. This property alone allows creating combinations of electronic orders that no naturally occurring material possesses. This possibility has given birth to the field designer quantum materials, where heterostructures are exploited to realize elusive quantum phases of matter not present in conventional compounds. 
In this review, we present a quantum materials cookbook point of view on how to achieve this and use three elusive quantum states engineered in vdW heterostructures as examples: topological superconductors, quantum spin-liquids and flat band systems.

The creation of topological superconductivity represents the first paradigmatic example of the possibilities brought by this flexibility. It is well known that topological superconductivity can be artificially engineered by combining s-wave superconductivity, spin-orbit effects, and magnetism. Materials with these properties can be combined in heterostructures of 2D materials by using layered superconductors, monolayer magnetic materials, and strong spin-orbit effects as the necessary ingredients of realizing topological superconductivity.

A second example consists of engineered quantum spin-liquids, highly entangled quantum magnets. The emergence of quantum spin-liquids is known to require a fine-tuning between spin-interactions, which is one of the limitations to finding them in non-tunable compounds. VdW heterostructures provide a way around this, with their possibility of finely tuning magnetic interactions in a two-dimensional magnet by a proper choice of 2D substrate. We will also discuss the prospects of realizing QSL in artificial systems. 

Finally, as the third example, we discuss how combining two-dimensional materials allow us to create dramatically new electronic dispersions beyond simple superposition of the electronic orders of parent compounds. The most dramatic case of this consists of the emergence of flat bands from a material with highly dispersive electrons. This is exemplified by structurally engineered on-surface graphene structures, and the whole family of twisted vdW heterostructures.


\section{Artificial vdW topological systems}
The engineering of novel topological states of matter\cite{RevModPhys.83.1057,RevModPhys.82.3045}
represents one of the milestones of current materials engineering.
While a variety of natural topological compounds have been identified in
nature\cite{Ando2013}, artificial engineered systems open new prospects for potential
technological applications with common compounds.
Ultimately, this topological engineering can ultimately
lead to the realization of states that no natural compound hosts.
Topological states of matter encompass a wealth of
states, including crystalline, higher-order and quasiperiodic
topological states. Here, we will focus on two paradigmatic
cases, namely quantum anomalous Hall insulators\cite{Liu2016} and topological
superconductors\cite{Alicea2012}. These two topological states represent critical
milestones for the fields of electronics and topological
quantum computing, respectively.

\subsection{Artificial topological superconductors}

\begin{figure*}
    \centering
    \includegraphics[width=0.98\textwidth]{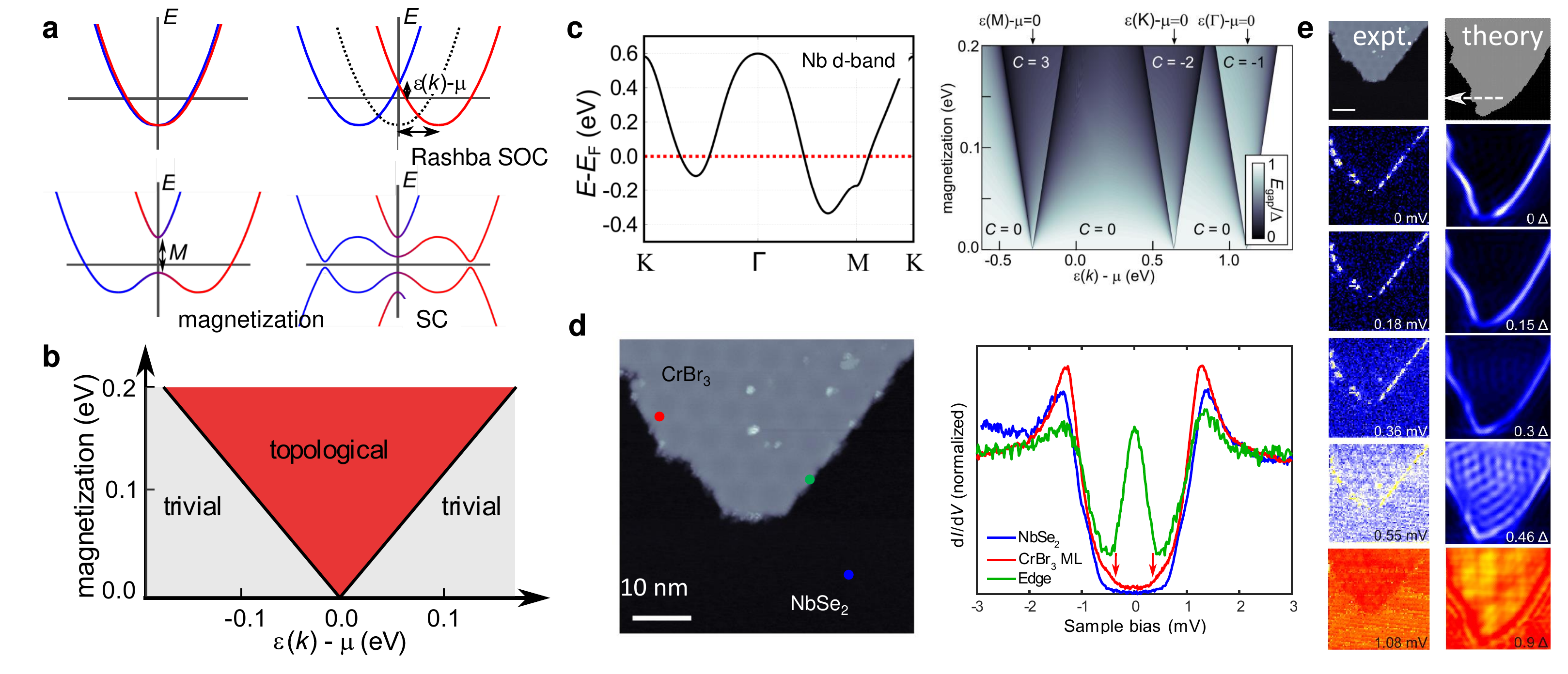}
    \caption{(a) Schematic of the band structure with the ``ingredients'' of topological SC: Rashba-type spin-orbit coupling, magnetization, and superconductivity. (b) Resulting topological phase diagram. (c) Nb d-band band structure in NbSe$_2$ and the topological phase diagram in this this system with a triangular lattice \cite{kezilebieke2020topological}. (d) Realization of the TSC in a van der Waals heterostructure. Left panel shows an STM topographic image of a CrBr$_3$/NbSe$_2$ heterostucture and d$I$/d$V$ spectroscopy on the three spots indicated is shown in the right panel \cite{kezilebieke2020topological}. (e) LDOS maps compared with theoretical modelling showing the presence of Majorana zero modes at the island edges \cite{kezilebieke2020topological}. }
    \label{fig:TSC}
\end{figure*}

The creation of topological superconductivity\cite{Kitaev2001} represents the first
paradigmatic example of the milestones of artificial engineering\cite{Alicea2012}. It 
is well known that topological superconductivity can be
artificially engineered by combining s-wave superconductivity, spin-orbit
effects, and magnetism\cite{Alicea2012,PhysRevLett.105.077001,PhysRevLett.105.177002,PhysRevLett.100.096407,PhysRevB.79.161408,Beenakker2013}. 
Materials with these properties can be combined in
heterostructures of 2D materials\cite{Geim2013} by using layered superconductors\cite{Ugeda2015}, monolayer
magnetic materials\cite{Huang2017,Gibertini2019}, and strong
spin-orbit effects as the necessary
ingredients to realize topological superconductivity.

Topological superconductivity represents one of the most pursued quantum states of
matter in modern condensed matter physics. Besides the interest 
in this state sparkled
from the rise of topological insulators, topological superconductors represent
one of the potential cornerstones for topological quantum computing\cite{Alicea2011,PhysRevX.6.031016,Frolov2020}.
However, topological superconductors are extremely rare 
in nature, and thus a great
amount of experimental efforts have been focused on engineering this state
in a variety of platforms\cite{Alicea2012,Beenakker2013}. The fundamental requirements 
for creating topological
superconductivity rely on creating an effective superconducting 
spin-triplet state\cite{Kitaev2001,RevModPhys.63.239}
starting from a conventional spin-singlet
s-wave superconductor. This can be achieved by creating 
fine-tuned spin textures
in materials combining exchange fields and strong Rashba spin-orbit coupling.
Based on this idea, a variety of proposals and realizations have been
demonstrated in the last years in semiconducting nanowires\cite{PhysRevLett.105.177002,Mourik2012,Lutchyn2018,Frolov2020}, atomically engineered chains \cite{PhysRevLett.111.186805,NadjPerge2014,Feldman2016,Ruby2017,Kim2018,Kamlapure2018,Steinbrecher2018,Schneider2020}
and topological insulators\cite{PhysRevLett.100.096407,PhysRevLett.116.257003}. 
In all these systems, the critical emphasis is put on combining
different materials containing magnetism and superconductivity, 
a task in which interface
physics is known to play a critical role\cite{PalacioMorales2019}.
Two-dimensional materials provide a unique opportunity in 
this direction, due to the
weak van der Waals forces that allow combining different layers\cite{Geim2013,Gibertini2019}, 
namely superconducting
and magnetic, on a single van der Waals heterostructure.

The requirement of these different order parameters to engineer a topological
superconductor can be easily rationalized.
In short, engineering topological 
superconductivity requires creating an effective
spinless superconductor, whose minimal
model gives rise to a topological superconducting state.
For this sake, let us start
with the simplest model for topological superconductivity: 
the one-dimensional Kitaev model\cite{Kitaev2001}.
This model considers spinless electrons on a one-dimensional
chain in the presence of a finite first nearest-neighbor pairing,
whose Hamiltonian takes the form

\begin{equation}
    H = \sum_n t c^\dagger_n c_n
    + 
    \sum_n \Delta c_n c_{n+1}
    + \text{h.c.}
    \label{eq:ts}
\end{equation}
It should be noted that for spinless fermions, on-site
superconductivity is forbidden from the 
fermionic anticommutation relations.
The previous Hamiltonian in Fourier space takes
the form
\begin{equation}
    H = \sum_n t \cos (ka) c^\dagger_k c_k
    + 
    \sum_n \Delta \sin (ka) c_k c_{-k}
    + \text{h.c.}
\end{equation}
giving rise to a fully gapped eigenspectra
$\epsilon_k = \sqrt{t^2 \cos{(ka)}^2 + \Delta^2 \sin {(ka)} ^2}$.
Despite its fully gapped structure, solving the Hamiltonian
Eq. \ref{eq:ts} with open boundary conditions gives rise
to a zero mode, which in the case of the
exactly solvable point $\Delta=t$ has
an associated eigenstate of the form 
$\gamma = \frac{1}{2} (c_0 + c^\dagger_0)$.
For finite chemical potential $\mu$,
an exponentially localized zero
mode exists, yet with a more complex
spatial profile.
In contrast with conventional fermions, this
creation operator is its own dagger $\gamma = \gamma^\dagger$.
This implies that these particles are their own antiparticles, which is expressed in this model through this mathematical property, as expected
from a Majorana operator. Similar models can be written
for a two-dimensional system, in which case the single-Majorana
mode becomes a propagating Majorana edge state
in an otherwise fully gapped spectrum. 

The central question of artificial topological superconductivity
is to find procedures of engineering an effective spinless
superconductor, starting
from spin-singlet superconducting term
of the form

\begin{equation}
H = \sum t_{ij} c^\dagger_i c_j
+     \sum_n \Delta c_{n,\uparrow} c_{n,\downarrow} 
+\text{h.c.}
\end{equation}
Many of the strategies to engineer topological superconductivity
rely on designing a pseudo-helical electron gas\cite{PhysRevLett.100.096407,PhysRevLett.105.177002} (states crossing the Fermi level have a spin that is locked to their momentum, i.e.~a certain momentum implies certain spin direction), yielding
an effective single degree of freedom and with a finite
projection on the spin-singlet state above,
which interestingly could be
directly engineered with two-dimensional
van der Waals topological insulators\cite{Zhao2020}.
The previous idea implies that
the electronic modes must have a finite spin-momentum coupling so that
the propagation direction depends on the spin channel. Such spin-momentum
coupling can be realized by different forms of spin-orbit coupling\cite{PhysRevLett.100.096407,PhysRevLett.105.177002}, or by
exploiting non-collinear magnetic textures\cite{PhysRevLett.111.186805,PhysRevLett.110.096403}. It is interesting to note that
these strategies work both in one and two-dimensions, and
as a result, recipes for one-dimensional topological superconductivity
can easily be generalized to two dimensions.

The typical recipe for achieving topological superconductivity
is illustrated in Fig.~\ref{fig:TSC}a. Starting with a parabolic band,
the addition of Rashba-type spin-orbit coupling and magnetization creates the
type of band structure required for TSC as explained above. The addition of
superconductivity completes the requirements and results in a system that
realizes the phase diagram shown in Fig.~\ref{fig:TSC}b. When the chemical
potential is tuned to the band crossing point at $k=0$, even a very small
magnetization is sufficient to drive the system into the topological phase.
If the chemical potential is tuned away from this point, then stronger
magnetization is needed. 
Although this procedure requires very precise
fine-tuning between the system parameters,
it has been successfully demonstrated
for a variety of semiconductor devices\cite{Mourik2012,Lutchyn2018},
and van der Waals materials\cite{kezilebieke2020topological}.
Finally, it is worth emphasizing that besides the Majorana edge modes,
topological superconductors are also expected to show
Majorana excitations at domain walls\cite{PhysRevLett.122.187001,Wang2020TSC} and vortices\cite{Wang2018,Zhang2018,Zhu2019,PhysRevX.8.041056}.

\subsection{Materials for artificial topological superconductors}
The section above lays out the rather stringent requirements for realizing topological superconductivity, and we need materials that will retain their magnetic and superconducting properties in a heterostructure. This strongly suggests using vdW materials: this allows for a rational design of the heterostructure as we expect to retain the intrinsic properties of the different constituents. Topological superconductivity has been realized in atomic-scale structures using conventional materials (e.g.~iron atom chains on a lead or rhenium substrates \cite{NadjPerge2014,Feldman2016,Ruby2017,Kim2018}, cobalt islands under a Pb monolayer, and iron islands on an oxygen-terminated rhenium substrate \cite{Menard2017,PalacioMorales2019}). However, considering the strong chemical bonding between the materials in this case, these systems are susceptible to disorder, and interface engineering might be required in some cases \cite{PalacioMorales2019}.

While many monolayer ferromagnet materials are available for exfoliation (e.g.~CrI$_3$), they are very reactive, and accessing the topological edge modes in scanning probe microscopy and other experiments requires the system to have very clean edges. This points out towards the use of e.g.~molecular-beam epitaxy (MBE) growth and luckily high-quality growth of several materials has been demonstrated (Fe$_3$GeTe$_2$ \cite{Liu2017}, CrBr$_3$ \cite{Chen2019,kezilebieke2020electronic}). For the superconductor material, typical suggestions would include the 2H phase of the NbS$_2$, NbSe$_2$, TaS$_2$, and TaSe$_2$ \cite{delaBarrera2018,Zhao2019}. The scheme for realizing TSC is also applicable to bulk superconducting substrates, there the magnetic layer will couple strongly to the top layer of the SC, and as long as the substrate has relatively weak interactions between the layers, it is expected to work similarly to the monolayer case \cite{Menard2015,kezilebieke2020topological}. These real materials have hexagonal symmetry, which is reflected in the band structure. Instead of a single high-symmetry point in the Brillouin zone, there are several ($\Gamma$, $M$, and $K$ points), and the topological superconducting phase can be realized at any of these points. This means that tuning the Fermi level across the relevant band (e.g.~the Nb d-band in the case of NbSe$_2$), there are three different topological phases that have different Chern numbers as illustrated in Fig.~\ref{fig:TSC}c. In a real vdW heterostructure, the doping of the substrate will determine whether the system will enter a topological phase. 

This route to TSC has been realized experimentally in CrBr$_3$ / NbSe$_2$ heterostructures \cite{kezilebieke2020topological,kezilebieke2020electronic}. As can be seen from the calculated Nb d-band bandstructure shown in Fig.~\ref{fig:TSC}c, the $M$ point is closest to the Fermi level, and it is likely that the topological phase arises from this point. Experimentally, the strongest signature is the Majorana edge modes that appear at the interface between the trivial and topological phases. This is shown in Fig.~\ref{fig:TSC}d, which shows an STM topographic image of CrBr$_3$ island on a bulk NbSe$_2$ substrate and three d$I$/d$V$ spectra (the signal is proportional to the local density of states, LDOS, at the position of the STM tip): on the NbSe$_2$ substrate (blue), on the CrBr$_3$ island (red) and right at the edge of the island (green). The spectrum recorded on the island edge has a strong peak centered around the Fermi level (zero bias) consistent with the expected LDOS corresponding to the Majorana zero modes.  Fig.~\ref{fig:TSC}e shows the measured (left) and theoretical LDOS (right) as a function of the energy. At the Fermi energy, both the bulk phases are gapped, and only the Majorana modes at the edges of the islands are visible. As the energy is increased, we eventually start to see excitations in the topological superconductor with the edge modes overlapping with bulk states. Finally, above the superconducting gap, all significant LDOS contrast is lost.

Comparison between theory and experiment allowed estimating the values of the model parameters, namely, the induced magnetization in the top NbSe$_2$ layer due to the proximity of the CrBr$_3$ layer and the magnitude of the Rashba spin-orbit coupling. These estimates suggest that the magnetization and the spin-orbit coupling are of a similar magnitude, a few tens of meV. This values were also consistent with density-functional theory (DFT) calculations and in-line with proximity induced exchange coupling in CrI$_3$/WSe$_2$ and CrBr$_3$/MoSe$_2$ heterostructures \cite{Zhong2020,PhysRevLett.124.197401}.
Finally, the moir\'e pattern between CrBr$_3$ and
NbSe$_2$ was suggested to further stabilize the topological
superconducting state\cite{2020arXiv201109760K}.

\subsection{Artificial Chern insulators}
\begin{figure*}
    \centering
    \includegraphics[width=0.95\textwidth]{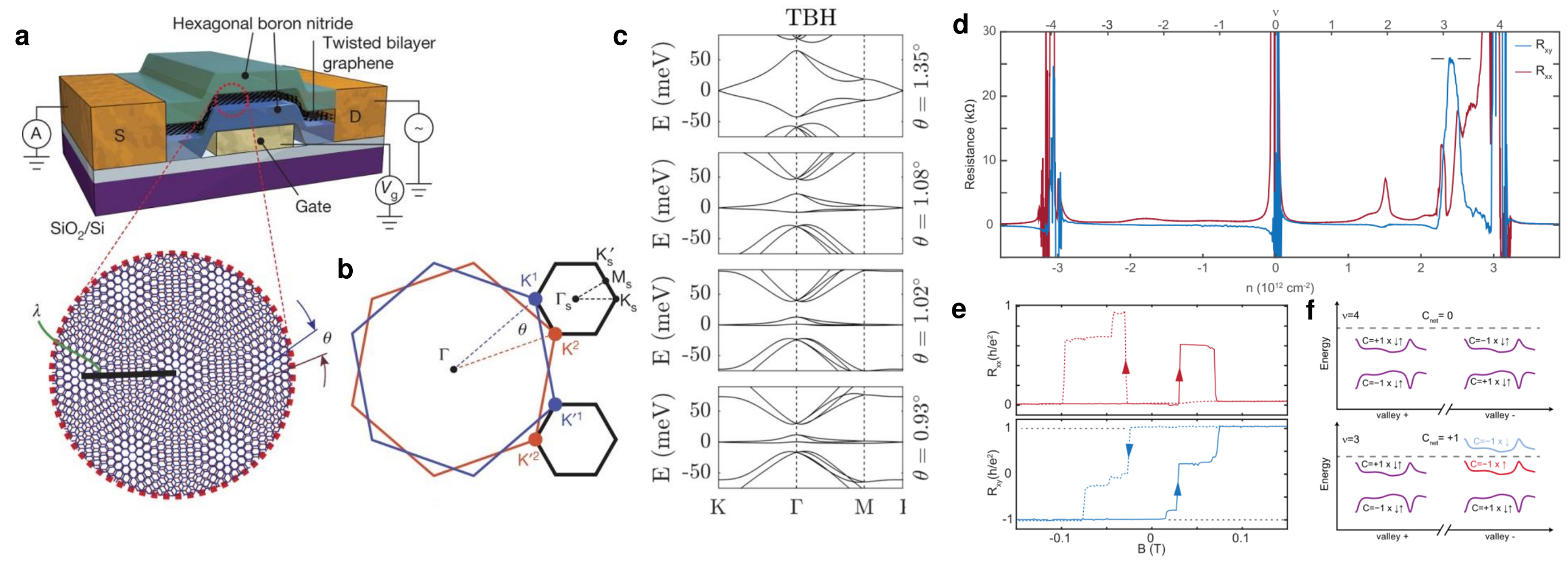}
    \caption{Chern insulating states in twisted bilayer graphene. (a) Illustration of a typical state-of-the-art device for TBG transport experiment. TBG is sandwiched between two layers of h-BN and atomically smooth graphite flake is used as the gate electrode. the contacts are so called edge contacts that have high transparency and avoid unwanted doping of the TBG \cite{Wang2013}. (b) The moir\'e modulation creates moir\'e mini-Brillouin zones at the $K$ and $K'$ points of the two graphene layers. These valleys are well-separated and each valley hosts two Dirac cones of the same chirality. (c) Evolution of the band structure as the twist angle is tuned. At the magic angle, a flat band with narrow band width that is well separated from other bands is formed. Panels a-c from Ref.~\cite{Cao2018_insl}. (d) Longitudinal ($R_{xx}$) and the Hall ($R_{xy}$) resistances measured as a function of the carrier density on a magic angle TBG device at $T=1.6$ K and under an external magnetic field of $B=+150$ mT. (e) $R_{xx}$ and $R_{xy}$ measured at the filling factor $\nu=3$ as a function of magnetic field $B$. (f) Schematic band structure at full filling of a moir\'e unit cell ($\nu = 4$) and $\nu = 3$. Panels e-f from Ref.~\cite{Serlin2019}.}
    \label{fig:TBG_Chernl}
\end{figure*}

Chern insulators \cite{PhysRevLett.61.2015} represent another paradigmatic state of matter in two-dimensional
systems. Besides their conventional engineering by combining spin-orbit coupling and
exchange fields \cite{Liu2016}, van der
Waals materials offer a novel approach to engineered Chern
insulators. 
This new approach to engineer Chern bands specifically exploits moire patterns
in twisted two-dimensional materials. 
The emergence of flat bands stems from
a non-abelian elastic gauge field
and will be further addressed in more
detail in section \ref{sec:flatgauge}. As illustrated in Fig.~\ref{fig:TBG_Chernl}a-c, the varying lattice registry in twisted bilayer graphene creates a long-wavelength moir\'e pattern. This moir\'e modulation creates moir\'e mini-Brillouin zones at the $K$ and $K'$ points of the two graphene layers. These valleys are well-separated, and each valley hosts two Dirac cones of the same chirality. The Dirac cones living at the $K^1$ and $K^2$ (and at $K'^1$ and $K'^2$) hybridize, and when the respective Dirac points are sufficiently close to each other in the $k$-space, this hybridization results in the formation of a flat band with narrow bandwidth that is well separated from other bands is formed as the twist angle $\theta$ is tuned across the magic angle (Fig.~\ref{fig:TBG_Chernl}c). This emergence of flat bands in twisted graphene bilayers is the starting point for realizing the Chern insulator states\cite{PhysRevB.82.121407,Bistritzer2011}. 

Flat bands in twisted bilayers can be interpreted as pseudo-Landau levels of an artificial
gauge field, generated by the modulated
stacking in the unit cell\cite{PhysRevLett.108.216802}. Close to charge neutrality,
these flat bands yield an 8-fold manifold, two-fold degeneracy coming from electron-hole states, 
two-fold coming from valley symmetry and two-fold coming from spin\cite{PhysRevLett.99.256802,PhysRevB.82.121407,Bistritzer2011,PhysRevX.8.031087}. 
As Landau levels\cite{PhysRevLett.108.216802,PhysRevB.99.155415}, each flat band is expected
to carry a non-trivial Chern number, analogous
to conventional Landau levels of quantum Hall states\cite{PhysRevLett.49.405}. However,
the original system is time-reversal symmetric, implying that flat bands stemming from
opposite valleys will carry opposite Chern numbers\cite{Vozmediano2010,PhysRevLett.108.216802}. 
This property suggests that if valley
symmetry is spontaneously broken, for example, due to electronic interactions,
twisted graphene bilayers become natural Chern insulators\cite{PhysRevResearch.1.033126}. 
The breaking of valley symmetry takes place when electronic interactions
create a spontaneous symmetry breaking, leading to a filling of just on the
the valley flat bands. A specific feature that must be taken
into account is that due to the existence of Dirac points in the
electronic structure\cite{PhysRevLett.99.256802,PhysRevB.82.121407,Bistritzer2011,PhysRevX.8.031087}, leading to the Chern insulator regime requires
to first opening a gap at the Dirac points\cite{PhysRevResearch.1.033126,PhysRevLett.124.166601,Sharpe2019}. This is done by
taking aligned hBN layers with the twisted bilayer that induce a small
symmetry breaking in the twisted bilayer
lifting the original Dirac points.
Ultimately,
in the presence of partial filling, this could lead to the emergence
of fractional Chern states \cite{PhysRevLett.124.106803}.

\subsection{Materials for artificial Chern insulators}
These predictions were realized in twisted graphene bilayers with the twist
angle ($\theta=1.15^\circ$) tuned to yield flat bands in the electronic
spectrum \cite{PhysRevB.82.121407,Bistritzer2011,Serlin2019}. The sample
fabrication followed the usual ``tear and stack'' process
\cite{Kim2016,Cao2018_insl,Cao2018_SC}, but TBG was aligned with the
underlying h-BN layer. The alignment with
BN turns out to be critical in lifting the low energy
Dirac points, allowing for the
emergence of a valley polarized state. These state-of-the-art devices
typically use TBG encapsulated by h-BN layers, and atomically smooth graphite
flake is used as the gate electrode (see Fig.~\ref{fig:TBG_Chernl}a).
Finally, the stack is electrically contacted using so-called edge contacts,
which have high transparency and avoid unwanted
doping of the TBG \cite{Wang2013}.
Fig.~\ref{fig:TBG_Chernl}d shows the longitudinal ($R_{xx}$) and the 
Hall ($R_{xy}$) resistances measured as a function of the carrier density 
on a magic angle TBG device at $T=1.6$ K and under an external magnetic 
field of $B=+150$ mT \cite{Serlin2019}. As expected for a quantum Hall 
state, $R_{xy}$ reaches $h/e^2$ and $R_{xx}$ approaches zero when the
electron density is tuned to filling factor $\nu = 3$ ($\nu$,
where $\nu$ is the 
number of the electrons in the flat band per moir\'e unit w.r.t. no 
external doping, i.e.~$\nu$ can have values between -4 and 4). 
The previous phenomena are the hallmarks of the QAHE state, and
most importantly it is retained in the absence of the field as shown in
Fig.~\ref{fig:TBG_Chernl} - The Hall resistivity is hysteretic (Fig.~\ref{fig:TBG_Chernl}e), with a
coercive field of several tens of millitesla. The Hall resistivity is
quantized ($R_{xy} = h/e^2$) and the longitudinal resistivity remains small
through zero external magnetic field, which demonstrates that the quantum
anomalous Hall state is stabilized by spontaneously broken time-reversal
symmetry.
In particular, this time-reversal symmetry breaking is purely associated
to the valley sector, where spontaneous
symmetry-breaking
leads just one of the valley filled
as sketched in Fig.~\ref{fig:TBG_Chernl}f.
Finally, it worth to note
that it is quite typical in the TBG experiments that the observed phenomena are
device-specific, with minor differences in the device parameters being
decisive which states are formed. For example, robust, thermally activated,
trivial insulator behavior and the QAH state can occur in very similar
devices.

Typically, the Chern number can be estimated from the value of the quantized
Hall conductance, but this measurement requires working on a transport setup, and it
would be extremely interesting to be able to somehow measure the Chern number
directly and independently. Precisely this was done by 
a scanning tunneling microscopy-based technique to directly
measure the Chern numbers of the different Chern insulating states
\cite{Nuckolls2020}. The topological gaps can be identified by measuring the
LDOS as a function of the electron density in the system (controlled through
external doping) at different external magnetic fields. If the electron
density at which gap opening and closing takes place depends on the external
magnetic field, the transition can be identified as a topological transition.
Tracking the electron densities at which these transitions happen as a
function of the magnetic field $B$, gives direct access to the associated
Chern number $C$ via $\frac{\mathrm{d}n}{\mathrm{d}B}=C/\Phi_0$, where
$\Phi_0$ is the magnetic flux quantum. In addition to a host of levels
arising from the zeroth Landau level at $\nu=0$ with Chern numbers
$C=0,\pm1,\pm2,\pm3,\pm4,\pm8,\pm12$, the authors observe a hierarchy of
correlated Chern insulating phases with Chern 
numbers $C=\pm1,\pm2,\pm3$ emerging as a function of magnetic field from the
different filling factors $\nu=\pm3,\pm2,\pm1$, respectively. All these
phases are stabilized by a magnetic field.

In addition to the example above, the Chern insulating state and the
quantum anomalous Hall effect has also been realized in rhombohedral
(ABC-stacked) graphene trilayers and twisted monolayer - bilayer graphene
samples \cite{Chen2020_Chern,Polshyn2020,Chen2020_Yankowitz} and recent experiments on magic-angle bilayers also suggest the possibility of realizing fractional Chern insulator states \cite{Wu2021}.

\begin{figure*}[t!]
\centering

    \includegraphics[width=.6\textwidth]{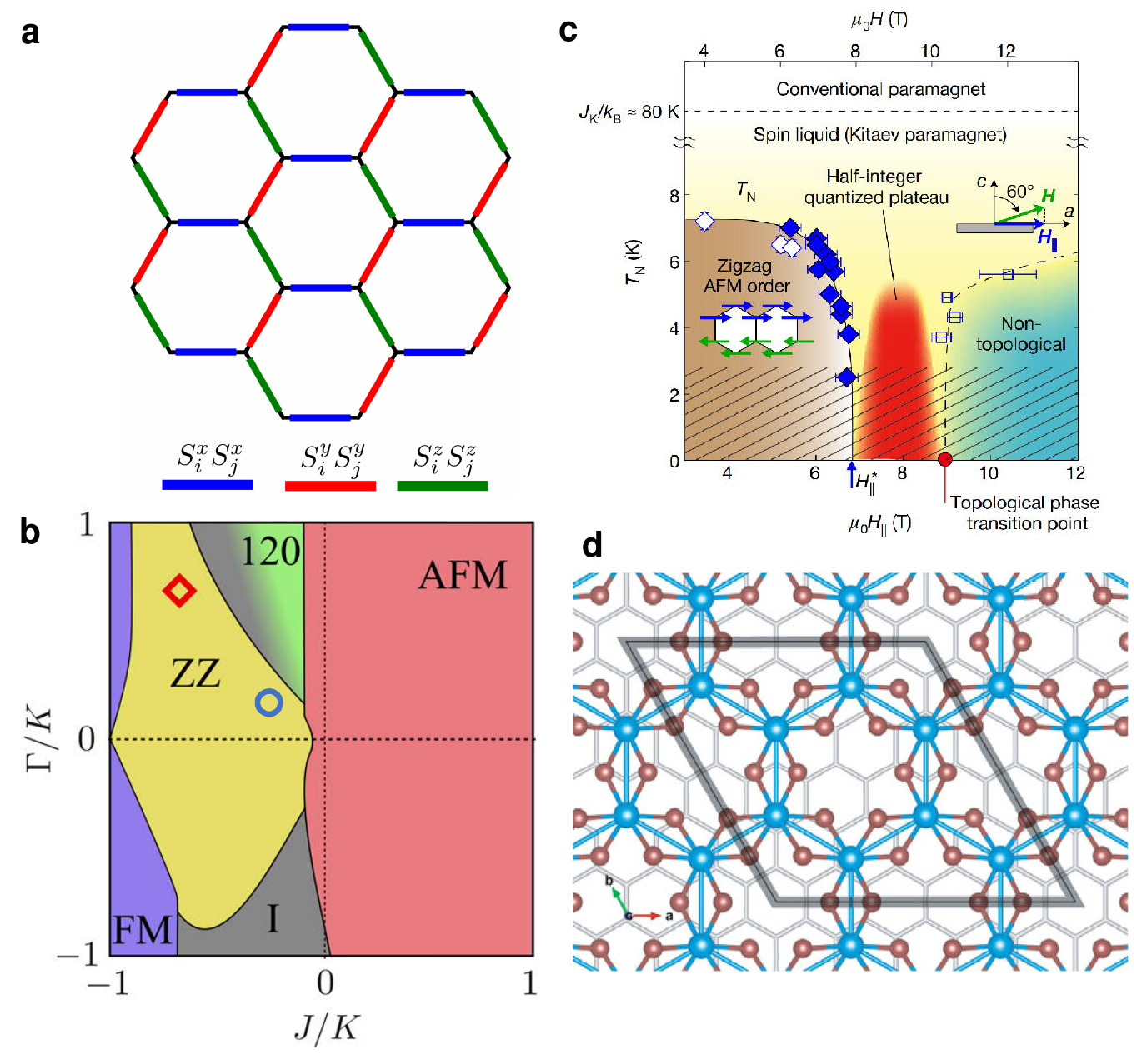}

	\caption{(a) Sketch of a spin model with frustration stemming from anisotropic interactions. (b) Phase diagram of RuCl$_3$ \cite{PhysRevB.91.241110}
	taking constant second and third neighbour
	exchange coupling. The red diamond indicates the estimated parameters for bulk RuCl$_3$ and the blue circle those for monolayer RuCl$_3$ on graphene \cite{PhysRevB.91.241110,PhysRevLett.123.237201,PhysRevLett.124.106804}. (c) The phase diagram of $\alpha-$RuCl$_3$ in an external magnetic field at $\theta = 60^\circ$. Open and closed diamonds represent the onset temperature of zigzag type AFM order based on thermal Hall conductivity measurements. A half-integer quantized plateau of the 2D thermal Hall conductance is observed in the red area and open blue squares represent the fields where the thermal Hall response disappears \cite{Kasahara2018}. (d)  Top view of $\alpha-$RuCl$_3$/graphene heterostructure, where the blue and maroon spheres represent Ru and Cl atoms, respectively, and the gray hexagon indicates the graphene monolayer \cite{PhysRevLett.123.237201}. }
\label{fig:qsl_kitaev}
\end{figure*}

\section{VdW quantum spin-liquids}

Quantum spin-liquids\cite{Anderson1973,Savary2016} are highly entangled quantum magnets, characterized by the emergence of
novel fractionalized particles. These many-body states are classified to their pattern
of long-range entanglement. In terms of their excitation spectrum, quantum spin-liquids
can be classified in gapped or gapless, and in a minimal picture, this is ascribed to
a gapless or gapped quasiparticle spinon excitations\cite{ANDERSON1987,Baskaran1987,Savary2016}. 
In addition, quantum spin-liquids can be described
by a deconfined gauge field theory \cite{Wegner1971,RevModPhys.51.659,Savary2016}, namely a theory that involves
free spinons capable of propagating \cite{PhysRevLett.63.322}. A fundamental question is, which mechanisms
capable of driving a system from a magnetically ordered state to a quantum disordered one can be designed.

\subsection{QSL from frustrated anisotropic interactions}
A first strategy to design quantum spin-liquids is to focus on models showing highly directional interactions that are inherently frustrated \cite{Kitaev2006}. This can
be done, for example, by taking square \cite{PhysRevLett.90.016803} 
or honeycomb lattices\cite{Kitaev2006}, and imposing
anisotropic spin-spin interactions that are dependent
on the bond considered.
As magnetic interactions in materials are
rotational symmetric in the absence of spin-orbit
coupling, these mechanisms are expected to
be realized in materials containing heavy atoms
in which spin-orbit effect compete and even overcome other interactions present \cite{PhysRevLett.105.027204,PhysRevLett.108.127203,Banerjee2016,Hermanns2018}.

Due to their interacting nature, the solution of quantum spin-liquid models represents one of the open problems in many-body physics. A great amount of insight can, however, be obtained from finely tuned models that allow for an exact solution. Among these specially tuned models, we encounter the Toric code and the anisotropic Kitaev honeycomb model\cite{Kitaev2006}. 
In particular, the Kitaev model realizes a highly anisotropic spin model in a  honeycomb
lattice that takes the form
\begin{equation}
    H = \sum_{\langle ij \rangle} S^\gamma_i S^\gamma_j
\end{equation}
where $\langle ij \rangle$ denote first neighbors and $\gamma$ labels the spin-component
that interacts for each bond as depicted in Fig.~\ref{fig:qsl_kitaev}a).
The genuine feature of the Kitaev honeycomb model stems from the possibility of obtaining an exact solution in terms of single particle excitations. Remarkably, the single-particle excitation are of Majorana type, and depending on the parameter regime, realize gapless or gapped Majorana states\cite{Kitaev2006}.

\subsection{Experiments QSL with anisotropic interactions}

Interestingly, the Kitaev honeycomb model \cite{Kitaev2006} 
(illustrated in Fig.~\ref{fig:qsl_kitaev}a) can be potentially adiabatically connected to quantum spin-liquid states realized in $\alpha$-RuCl$_3$ (RuCl$_3$) \cite{PhysRevB.91.241110,PhysRevLett.119.037201,Banerjee2016,Banerjee2017,Do2017}, and thus in the following we will focus on this compound. RuCl$_3$ is a layered Mott insulator with significant spin-orbit interactions that is in the close proximity to the  quantum spin-liquid ground state \cite{PhysRevB.91.241110,PhysRevLett.123.237201,PhysRevLett.124.106803}.
However, these materials often host complex Hamiltonians having several
contributions beyond the Kitaev exchange, including
first, second and third neighbor exchange, and
symmetric off-diagonal exchange\cite{PhysRevB.91.241110,PhysRevB.93.214431,PhysRevLett.120.077203,Winter2017}.
The model typically employed for this compound gives rise to the phase diagram sketched in Fig.~\ref{fig:qsl_kitaev}b
as a function of the first neighbor couplings,
keeping the second and third neighbor exchange
finite\cite{PhysRevB.91.241110}. 
In Fig.~\ref{fig:qsl_kitaev}b, the x-axis represents the ratio of the
Heisenberg ($J$) to Kitaev -type ($K$) spin coupling and $\Gamma$ is
symmetric off-diagonal exchange coupling. 
The whole diagram has been evaluated with the ratio of Hund's coupling
($J_\mathrm{H}$) to the Coulomb on-site interaction ($U$) of
$J_\mathrm{H}/U=0.2$, which can be estimated from ab initio calculations. 
It can be seen that the phase diagram hosts ordered magnetic phases ranging from ferromagnetic (FM) and antiferromagnetic (AFM) to more complicated zigzag (ZZ), 120 and incommensurate order (I) phases,
even without considering variations in the
further neighbor exchange. The best estimate for the parameters corresponding to bulk RuCl$_3$ is shown as a red diamond. This implies that the ground state of RuCl$_3$ is actually an ordered magnetic phase, which has been experimentally confirmed using, e.g., thermal Hall conductance measurements\cite{Kasahara2018}.

\begin{figure*}[t!]
    \centering
    \includegraphics[width=0.9\textwidth]{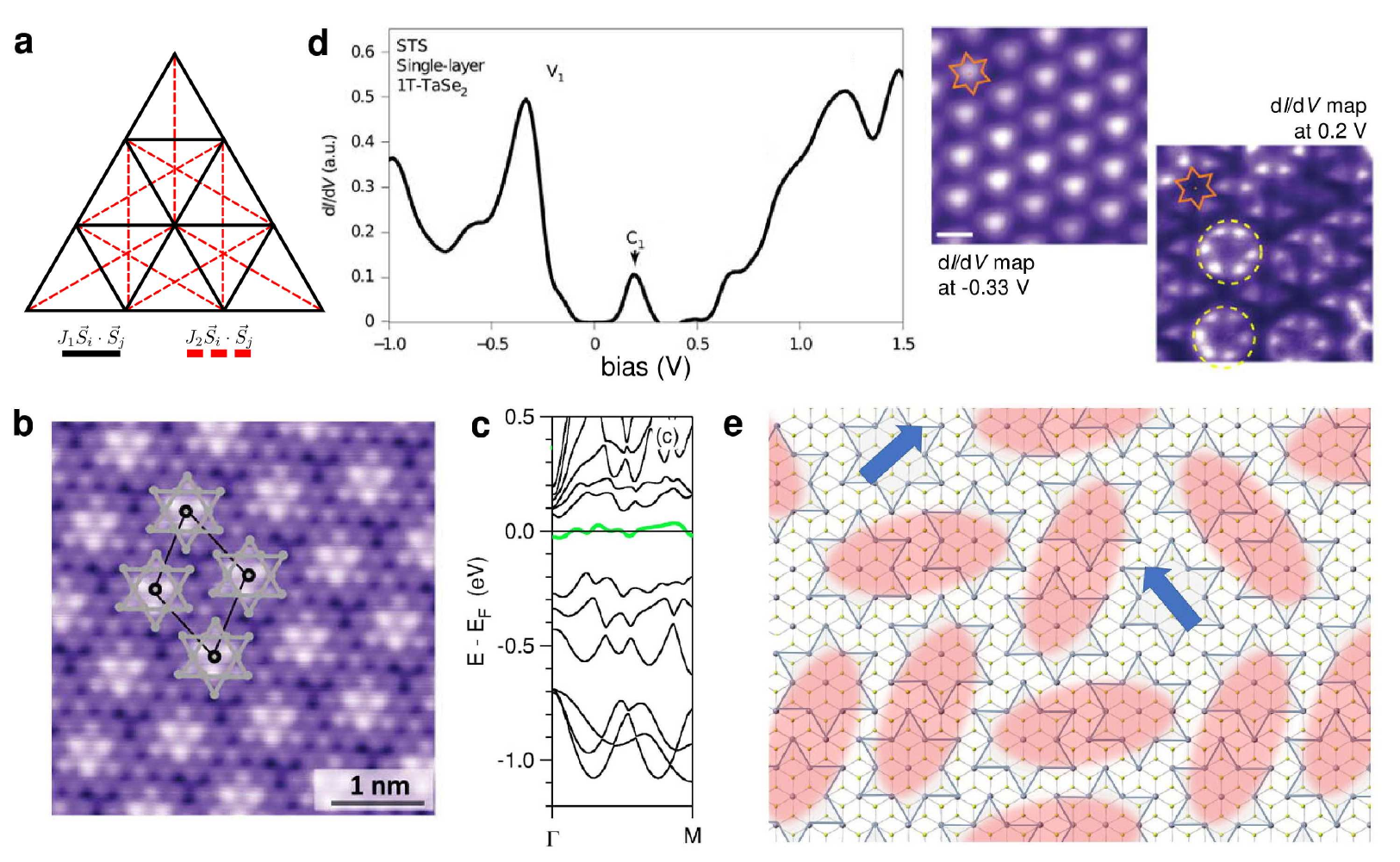}
    \caption{(a) Schematic of a quantum-spin-liquid with frustration stemming from geometric frustration. (b) STM image recorded on a buk 1T-TaS$_2$ surface showing the characteristic ``Star-of-David'' charge-density wave pattern \cite{TaS2_Mottness}. (b) Calculated band structure of 1T-TaS$_2$ with the $\sqrt{13}\times\sqrt{13}$ ``Star-of-David'' reconstruction and including the Ta spin-orbit coupling reveals a single distinct split-off banb at the Fermi level (green line) \cite{Rossnagel2006}. Including the Hubbard U-term splits this band into a lower and upper Hubbard band 
    (d) STM d$I$/d$V$ measurements on a related compound 1T-TaSe$_2$ monolayer  showing LHB and UHB \cite{Chen2020}. The panels on the right show constant height d$I$/d$V$ maps that allow direct visualization of the LHB and UHB wavefunction symmetries. (e) Illustration of the low-temperature state of 1T-TaS$_2$ with spatially random pairs of ``Star-of-Davids'' forming singlets. It is possible to form gapless, low-energy fractional excitations (blue arrows) \cite{Klanjek2017}. }
    \label{fig:QSL_geom}
\end{figure*}

Remarkably, it was experimentally demonstrated that applying a sufficient in-plane magnetic field can destroy the long-range order of the magnetic ground state and give rise to quantum spin-liquid behaviour \cite{Kasahara2018}. This is illustrated in the phase diagram shown in Fig.~\ref{fig:qsl_kitaev}c, where the boundaries of the different phases have been followed using thermal Hall conductance measurements.
Interestingly, the Majorana edge modes that arise in the quantum spin-liquid ground state can be directly verified as half-integer quantized thermal conductance, which is observed in the region shaded with red in Fig.~\ref{fig:qsl_kitaev}c. Further increasing the lateral magnetic field gives rise to a phase transition to some other non-topological phase.

In addition to the application of a lateral magnetic field, many other routes (e.g.~external pressure and chemical doping) are being tested to suppress magnetism, enhance the pure Kitaev interactions and drive the system towards the quantum liquid state. In the spirit of the designer material principles, we highlight a couple of theoretical ideas where heterostructures could be used to promote the quantum spin-liquid state. It has been proposed that monolayer RuCl$_3$ on graphene (illustrated in Fig.~\ref{fig:qsl_kitaev}d) would result in a system with enhanced Kitaev type interactions \cite{PhysRevLett.123.237201,PhysRevLett.124.106804}. By using ab initio calculations, it was shown that the RuCl$_3$ becomes strained and doped in this heterostructure. This might even drive an insulator-to-metal transition and help to realize predicted, exotic superconducting states in quantum spin-liquids\cite{PhysRevB.85.140510,PhysRevLett.123.237201}. In any case, the strain and doping are predicted to enhance the Kitaev interactions (increasing $K$ and decreasing $J$ and $\Gamma$) and move the system closer to the Kitaev. RuCl$_3$/graphene heterostructures have been also realized experimentally\cite{Mashhadi2019,PhysRevB.100.165426}, but not yet down to the monolayer limit. However, experiment on thicker RuCl$_3$ layers already give indications of the charge transfer and hybridization between the RuCl$_3$ and graphene bands\cite{Mashhadi2019,PhysRevB.100.165426}.

\subsection{QSL from geometric frustration}
Geometric frustration can lead to spin-liquid behavior and considering a simple picture of three spins can give a flavor of the general idea: Consider three spins at the corners of a triangle with antiferromagnetic interactions. This system does not have a configuration where all antiferromagnetic interactions can be simultaneously satisfied, i.e.~the system is frustrated. 
This geometric frustration leads to unusually large ground state degeneracies, already at the classical level. The situation described above corresponds to a classically frustrated system, in which
quantum entanglement between sites is not considered and corresponds to the
so-called spin ice models.
In the quantum realm, an effective strategy to realize quantum spin-liquid physics is to focus
on models realizing non-bipartite lattices, such as triangular and kagome lattices. Kagome lattice
models\cite{Yan2011,Han2012,PhysRevLett.109.067201} have been known to be a paradigmatic platform for quantum
spin-liquid physics. Since triangular
lattices are often more common in the van der Waals world, we will in the
following focus on that case.
Focusing on the triangular lattice model,
in the minimal case in which only first neighbor interactions are considered, the ground state is actually an ordered
state with 120 degrees spin spiral\cite{PhysRevB.45.12377,PhysRevLett.82.3899,PhysRevLett.99.127004}. However, this model 
can be pushed to a more frustrated regime by including
additional interactions\cite{PhysRevLett.120.207203,PhysRevX.9.021017}, and
in particular, a second neighbor exchange coupling\cite{PhysRevLett.123.207203}, driving
the system to a quantum spin-liquid ground state.
Although an exact solution cannot be obtained in this limit, tensor network calculations have shown strong signatures of a gapless QSL liquid state in this regime, featuring gapless Dirac spinons.\cite{PhysRevLett.123.207203}

The low energy excitations of these models 
in terms of chargeless emergent fractionalized
excitations with $S=1/2$ known as spinons.
We start with a Heisenberg model of the form
$
    H = \sum_{ij} J_{ij} \vec S_i \cdot \vec S_j
    \label{eq:heisenberg}
$
where $S_i$ are the local spin operators. 
Assuming a quantum spin-liquid ground state,
we can express the localized
spins as emergent chargeless
$S=1/2$ fermions of the form 
$S^\alpha_i = \sum_{s,s'} \sigma^\alpha_{s,s'}  f^\dagger_{i,s} f_{i,s'}$,
where $f^\dagger_{i,s}$ denotes the creation operator of a fermionic spinon
in site $i$,
and $\sigma^\alpha_{s,s'}$ are the spin Pauli matrices. 
The localized moment is implemented by enforcing having 
a single fermion in each site $f^\dagger f = 1$.
At the mean field level,
the Heisenberg Hamiltonian
a tight binding model of
free propagating $S=1/2$ spinons
of the form

\begin{equation}
    H = \sum_{ij} \chi_{ij} f^\dagger_{i,s} f_{j,s'}
\end{equation}
where $\chi_{ij}$ are the mean-field parameters of the 
mean-field Hamiltonian.
The spinon excitations of the quantum spin-liquid state 
can thus be understood
from the spinon dispersion. For example, 
gapless Dirac quantum spin-liquids
have an associated spinon model featuring Dirac points,\cite{PhysRevLett.123.207203} 
whereas models with a finite
spinon Fermi surface are stem from 
model with a finite 
Fermi surface. This classification is often used when
characterizing quantum spin-liquid ground states,
and has direct impact on the temperature-dependence of the
thermal conductivity\cite{2020arXiv200715905M}.

\subsection{Experiments on geometrically frustrated QSLs}

The necessary ingredients for a QSL - triangular lattice with frustrated magnetism (Fig.~\ref{fig:QSL_geom}a) - can be realized in van der Waals materials. This has been demonstrated in the 1T phase of TaS$_2$ (1T-TaSe$_2$ is expected to be similar), where the presence of various charge-density wave (CDW) states (depending on the temperature) has been known for some time \cite{Wu1989,Wu1990,Park2019}. The low-temperature CDW state results in a $\sqrt13\times\sqrt13$ reconstruction of the 1T-TaS$_2$ lattice that has a 13 Ta atom ``star of David'' unit cell \cite{Wu1989,Wu1990,Rossnagel2006,TaS2_Mottness,Park2019} as illustrated in Fig.~\ref{fig:QSL_geom}b. This causes folding of the band structure and, together with modified hoppings caused by the reconstruction and the presence of spin-orbit coupling, results in a single band with a relatively flat dispersion at the Fermi level (Fig.~\ref{fig:QSL_geom}c) \cite{Rossnagel2006}. In the presence of strong electron-electron interactions ($U$ larger than the bandwidth of band at Fermi level), the system will undergo a Mott metal-insulator transition and instead of the single band at the Fermi level, there will be a fully occupied lower Hubbard band (LHB) below the Fermi level and a fully unoccupied upper Hubbard band (UHB) above it \cite{Law6996}.

In the case of 1T-TaS$_2$, this Hubbard band correspond to a single unpaired electron per CDW ``star of David'' unit cell, which are the building block of the quantum spin-liquid state in this material. The Hubbard bands have been demonstrated in bulk 1T-TaS$_2$ \cite{PhysRevB.92.085132,TaS2_Mottness} and also in monolayer 1T-TaSe$_2$ \cite{Chen2020}. As illustrated for 1T-TaSe$_2$ in Fig.~\ref{fig:QSL_geom}d, tunneling spectroscopy allows direct verification that the system is gapped and the energies of the LHB and UHB can be easily probed. In addition, by mapping the spatial variation of the tunneling conductance d$I$/d$V \propto$ LDOS, the spatial symmetries of the states can be probed. In the case of 1T-TaSe$_2$ monolayer, it can be seen that the orbital texture of LHB and UHB are different (right side of Fig.~\ref{fig:QSL_geom}d).

\begin{figure*}[t!]
    \centering
    \includegraphics[width=2\columnwidth]{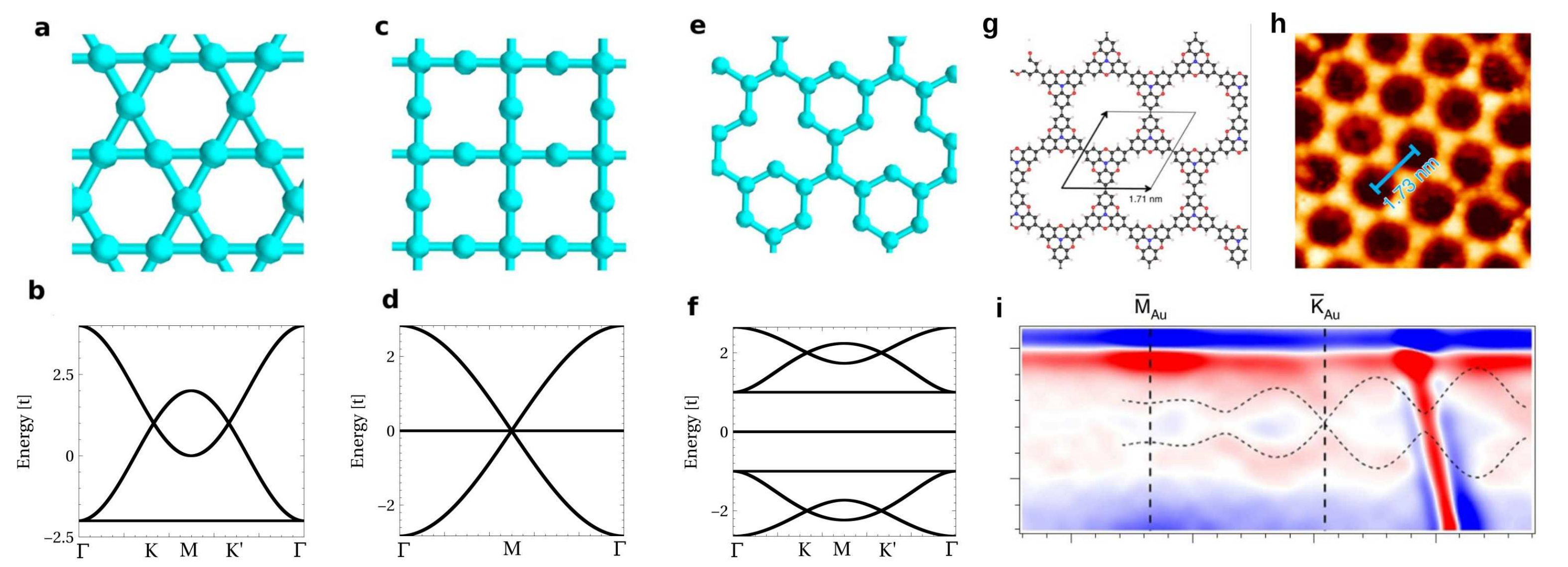}
    \caption{Structure (a,c,e) and band structure (b,d,f)
    of different lattices featuring flat bands:
    the kagome lattice (a,b), the Lieb lattice (c,d) and
    a modified honeycomb lattice (e,f). (g,h) Schematic and experimental structure of a $\pi$-conjugated polymer poly-trioxaazatriangulene \cite{Galeotti2020}. (i) Second-derivative ARPES signal of the sample shown in panel h acquired at a photon energy of 120 eV. The dashed lines represent calculated bands for free-standing polymer netowrk, shifted downwards by 0.12 eV \cite{Galeotti2020}.
    }
    \label{fig:flat_lattice}
\end{figure*}

While STM and tunneling spectroscopy can be used to probe the Hubbard bands, it is difficult to directly use these techniques to probe the spin-liquid state. This is usually done with neutron scattering, where the ``smoking gun'' for the QSL state is the lack of magnetic order down to the lowest temperatures. The other option is muon-spin-relaxation, which has been applied to bulk 1T-TaS$_2$ to show that the spin excitations are gapless, and there is no long-range order in temperatures of at least down to 70 mK \cite{Klanjek2017}. Those experiments show that below 55 K, there is a broad distribution of the relaxation times indicating a highly inhomogeneous magnetic phase at all Ta sites. This is strong evidence that there is growing randomness in the spin system as temperature decreases below 55 K. The observed slowing down of spin fluctuations is consistent with the freezing of singlets as illustrated in Fig.~\ref{fig:QSL_geom}e. Interestingly, for 1T-TaS$_2$ the resonant valence bonds are formed
between magnetic moments with an extension of the enlarged
unit cell generated by the CDW, in
comparison with the atomic-like moments of 
bulk QSL candidates.

The problem with the bulk probes such as neutron scattering or muon spin-relaxation is that they are typically not sufficiently sensitive to probe monolayer samples. There are theoretical suggestions that tunneling spectroscopy could be used for this even though the magnitude of the predicted signal would depend on the measurement geometry (e.g. 2D junction vs. STM) and the type of the spin-liquid \cite{Konig2020_QSL}. In addition,
and despite their chargeless nature,
signatures of spinon interference can be potentially
probed by inelastic transport spectroscopy\cite{ruan2020imaging}. 
Finally, muon spin-relaxation has been used to probe the Kondo 
effect with spinons\cite{Gomilek2019},
by probing the existence of a spinon-Kondo
cloud around magnetic impurities, and this technique could perhaps be extended to monolayer samples.

\section{New VdW flat bands}

The engineering of flat bands has been at the forefront of condensed matter
physics for a long time. Flat band systems are characterized by having
almost dispersionless states, which in the presence of any residual
interactions are expected to be prone to a variety of electronic
instabilities \cite{Leykam2018,PhysRevLett.62.1201,PhysRevB.83.220503}. In the following, we will discuss several directions
that van der Waals materials provide towards the realization of flat
band systems.

\subsection{Generating flat bands from geometric frustration}
The simplest instance in which flat bands appear in electronic systems
are tailored lattices leading to destructive interference\cite{Leykam2018,PhysRevB.78.125104}. Paradigmatic
examples of these flat band models are Lieb and kagome lattices\cite{PhysRevB.78.125104}. In this system, electron propagation is quenched due to the existence of
complementary paths that interfere destructively. 
This destructive interference
can be often weakened
by adding additional perturbations.
For example, next nearest neighbour (NNN) hoppings interactions will often cause the flat bands to acquire dispersion, as
flat bands are localized eigenstates on ``disconnected'' lattice
sites and NNN hoppings connect these sites and make the flat band dispersive. In the case of the Lieb
lattice, the existence of a flat band can also be understood from Lieb's
theorem\cite{PhysRevLett.62.1201}.
In its general form, this theorem states that
for a fully bipartite lattice,
the number of flat bands will
be $|N_A-N_B|$,\cite{PhysRevLett.62.1201}
where $N_A$ is the number of removed
sites from sublattice $A$, and
$N_B$ the number of sites removed
of sublattice $B$.
In particular, this implies that
generic bipartite lattices in which one site is removed will show a
flat band. The Lieb lattice can be built by removing one
site from the square lattice, leading to the existence of a flat band.
In this very same fashion, other flat band models can be systematically
constructed by removing a certain number of sites.
These types of flat bands have been realized in artificial systems based on atomic lattices \cite{Drost2017,Slot2017,PhysRevResearch.2.043426,Khajetoorians2019,Yan2019}. They can also be formed in suitable engineered, chemically synthesized lattices, where covalent organic frameworks and metal-organic frameworks are especially attractive systems for realizing these
artificial models \cite{Wang2013TI,Wang2013TI2,Dong2016,Zhang2016,PhysRevLett.116.096601,Kumar2017,Sun2018,Springer2020,Pawlak2021}.

\subsection{Experiments flat bands from frustration}

There have been many theoretical proposals on metal-organic frameworks (MOFs) with kagome structure that should result in flat bands in their band structure \cite{Wang2013TI,Wang2013TI2,Zhang2016}. However, the experimental demonstration has proven difficult. If the assembly is carried out directly on a metal substrate (typically Au(111), Ag(111) or Cu(111)), it is relatively straightforward to realize MOFs with a large degree of structural perfection \cite{Dong2016}. Unfortunately, the relatively strong interaction with the underlying metal substrate typically masks the intrinsic electronic structure of the MOF. On the other hand, the formation of the high-quality MOFs on weakly interacting substrates is much more challenging \cite{Kumar2018,yan2020synthesis} and unambiguous proof of the flat bands has not been demonstrated.

The other chemical strategy for synthesizing two-dimensional networks on surfaces relies on the formation of covalent carbon-carbon bonds and structures called covalent organic frameworks (COFs). While there are extensive results on the formation of the one-dimensional nanocarbons (graphene nanoribbons) \cite{Cai2010,Ruffieux2016,Talirz2016,Grning2018,Rizzo2018,Yan2019}, challenges remain to create two-dimensional assemblies with very high quality \cite{onsurface2018,Clair2019,Grill2020}. However, there are recent experimental results that are pushing this field towards higher quality samples towards the formation of flat bands in the MOF or COF band structure \cite{Rizzo2020,Galeotti2020,Pawlak2021,yan2020synthesis}. The realized strategies rely on making a molecular network with a kagome lattice with one of the examples highlighted in Fig.~\ref{fig:flat_lattice}g-i. In particular, on-surface polymerization was used to realize a high-quality two-dimensional polymer poly-trioxaazatriangulene network \cite{Galeotti2020}. This sample was of sufficiently high quality to allow angle-resolved photoemission spectroscopy (ARPES) experiments that can be used to directly probe the structure of the occupied bands as shown in Fig.~\ref{fig:flat_lattice}i. This shows the folded bands of the valence band of the polymer that match the expected results well (calculated bands shown by dotted lines). The kagome flat band is expected to be at the bottom of the conduction band and cannot be directly probed by ARPES experiments. These results are along the path towards tuneable 2D organic or metal-organic structures with engineered flat bands. The incorporation of metal atoms with magnetism or a large spin-orbit interaction opens additional possibilities in realizing topological materials \cite{Wang2013TI,Wang2013TI2,Yan2018,Gao2019}.

\subsection{Generating flat bands from long wavelength modulations}
A simple way of generating nearly flat bands consists of weakly
coupling quantum dot states. In this picture, the bandwidth is determined by the coupling between the quantum dots - the weaker it is, the flatter the resulting bands will be. A convenient way of achieving this in a real material in a large 
scale is by exploiting moir\'e patterns \cite{PhysRevLett.121.266401,PhysRevB.99.235417,PhysRevLett.124.206101,PhysRevB.98.224102,Weston2020,Xian2019,PhysRevB.102.075413}. The fundamental idea relies on the locally modulated stacking over the moir\'e pattern that causes a spatial modulation of the conduction and valence band edges and leads the formation of a large scale array of quantum dots in twisted van der Waals superlattices. The mechanism for flat band generation can be rationalized from the decoupled limit, in which the system consists of decoupled quantum dots. The twist angle between the layers changes the size and separation between the quantum dots, promoting a finite hybridization between them that leads to nearly flat bands\cite{Xian2019,2020arXiv200914224Z,2020arXiv200801735A}. It is worth noting that this mechanism holds when there is a bandgap in the original materials (e.g.~twisted h-BN and twisted dichalcogenide systems). This mechanism also requires the existence of a confinement gap. As a result, semimetals like graphene, in which electrons cannot be electrostatically confined, require a different mechanism for flat band generation. We will illustrate the use of gauge fields for this in section \ref{sec:flatgauge}.

\subsection{Experiments flat bands from quantum dots}
As we discussed above, flat bands can be realized in gapped, twisted moir\'e systems and this has been demonstrated in several experiments. An early experiment by Zhang et al.~relied on direct growth of rotationally aligned MoS$_2$/WSe$_2$ heterostructure, where the lattice mismatch then creates a moir\'e pattern \cite{Zhang2017}. While not directly resolving the flat bands spectroscopically, they demonstrated that the system had the necessary ingredients for their existence: the modulated  interlayer coupling giving rise to a modulation of the conduction and valence band edge energies. They showed that the valence and conduction band edges are located at different layers and that the local bandgap was periodically modulated with an amplitude of $\sim0.15$ eV, leading to the formation of a two-dimensional electronic superlattice.

The flat bands were directly identified in a later study \cite{Zhang2020}, which concentrated on a twisted bilayer WSe$_2$ samples with twist angles of $3^\circ$ and $57.5^\circ$. By using scanning tunnelling spectroscopy, it was possible to directly map the spatial extent of the wavefunctions at the flat-band energy and to show that the localization of the flat bands depends on the twist angle. The observed flat bands originated from the highest valence band at the $\Gamma$ point (the conduction band onset varies very little over the moir\'e pattern and hence does not result in the type of quantum dot states required for the formation of the flat bands). The flat band in $3^\circ$ twisted bilayer is localized on the hexagonal network separating the AA sites where as in the $57.5^\circ$ systems, it is localized on the AB sites. These observations match well with the results of earlier atomistic calculations \cite{PhysRevLett.121.266401}.

While the basic physics of these systems can be understood with only considering the spatially varying stacking, in real materials, additional effects are expected to take place. For example, it is likely that there are some atomic-scale structural relaxations over the moir\'e pattern. This is precisely the effect that was assessed in the paper by Li et al.~\cite{Li2021}, who focussed on the twisted WSe$_2$/WS$_2$ system and used a combination of scanning tunneling spectroscopy (STS) experiments and ab initio simulations of TMD moir\'e superlattices. They find a strong 3D buckling reconstruction together with large in-plane strain redistribution in their heterostructures. Using STS imaging, they identify different types of flat bands originating either from the $K$-point at the valence band edge or from the $\Gamma$-point that gives rise to more deep-lying moir\'e flat bands. By analyzing the origin of these flat bands in detail, it is revealed that the $K$-point flat bands are mainly a result of the deformation of the monolayer. Similar behavior can be reproduced by considering only a puckered monolayer WSe$_2$. On the other hand, the $\Gamma$-point flat bands are more in-line with the idea of the moir\'e induced, weakly coupled array of quantum dots. We will discuss the effects of periodic strain in more detail in section \ref{sec:flatgauge}.

The flat bands in the twisted TMD bilayers where the electron kinetic energy is suppressed are of course, fertile ground for realizing systems where interactions play a dominant role. There have been several publications on e.g.~realizing different kinds of correlated states, correlated insulators and Wigner crystals in WSe$_2$/WSe$_2$ and WSe$_2$/WS$_2$ moir\'e superlattices \cite{Wang2020,Tang2020,Regan2020}. However, the moir\'e flat band systems can also have exciting optical effects and this has given birth to a field studying moir\'e excitons \cite{Seyler2019,Tran2019,Jin2019,Alexeev2019}.

When the moir\'e period is larger than the exciton Bohr radius (around $\sim1-2$ nm in e.g. MoSe$_2$ and WSe$_2$), the excitons will experience a spatially modulated periodic potential from the moir\'e. The other design parameter in heterobilayers is the relative alignment of the conduction and valence band edges, which allows the formation of intralayer excitons (e.g.~WSe$_2$/WS$_2$ system where the electron and the hole reside in the same layer \cite{Jin2019}), interlayer excitons (e.g.~MoSe$_2$/WSe$_2$ system where the electron and the hole exist in different layers \cite{Seyler2019,Tran2019}) and hybridized excitons (e.g.~MoSe$_2$/WS$_2$ where the electron (for this system) is delocalized in the two layers \cite{Alexeev2019}). Finally, the moir\'e-defined quantum dots preserve the three-fold rotational (C3) symmetry, which implies that e.g.~the  interlayer excitons should inherit valley-contrasting properties \cite{Seyler2019}. These systems are currently under intense study to realize arrays of entangled quantum light emitters and realizing new exotic excitonic many-body phases (e.g.~topological exciton insulator) \cite{Wang2019,Tartakovskii2019}. 

\begin{figure*}[t!]
    \centering
    \includegraphics[width=\textwidth]{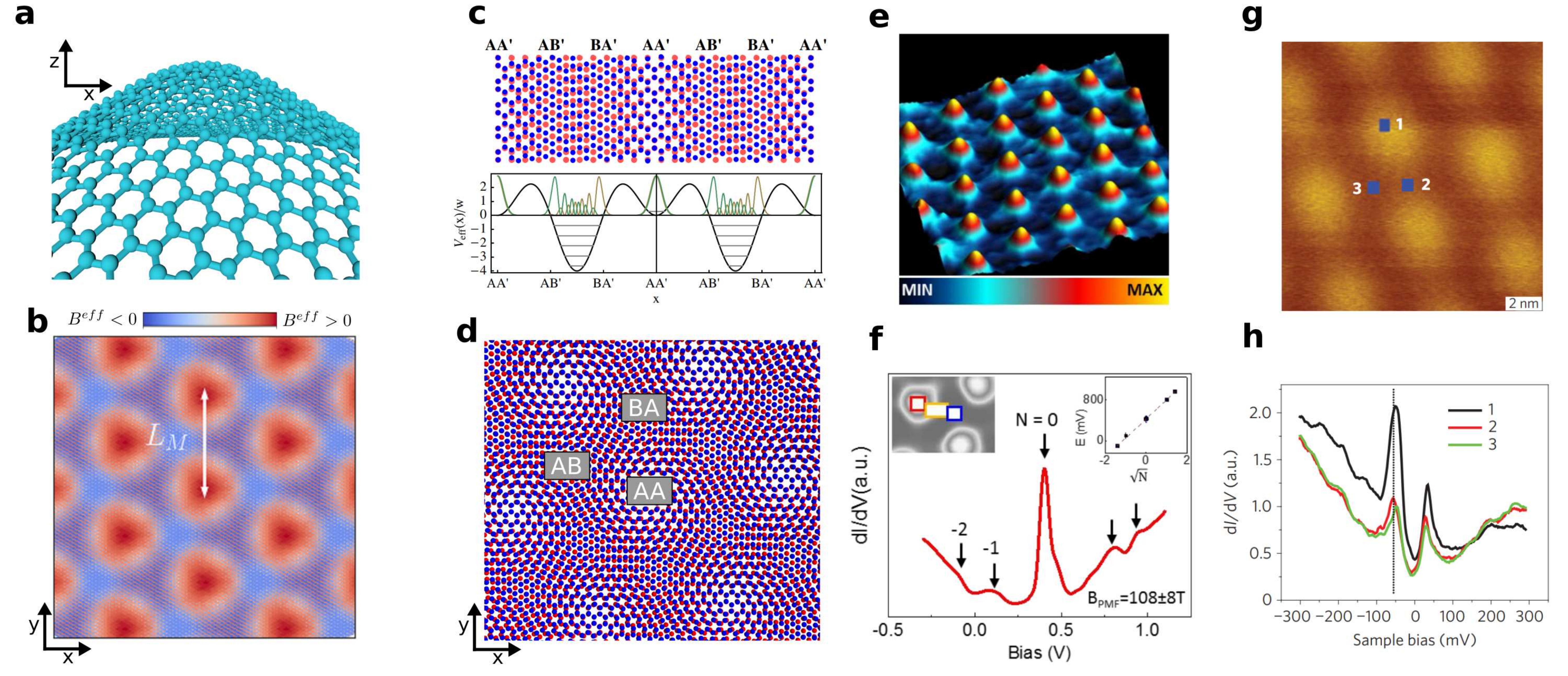}
    \caption{Sketch of the buckling of a graphene monolayer
    (a) and spatial profile of the induced gauge field (b) \cite{Manesco2020}.
    Change in the local structure of a graphene bilayer
    leading to a non-abelian gauge field (c) \cite{PhysRevLett.108.216802},
    and profile of the stacking in the moire unit cell (d).
    Panels (ef) shows the topography (e) and
    dI/dV (f) of buckled graphene
    monolayer, showing the emergence
    of pseudo Landau levels.\cite{Mao2020}
    Panel (f,g) shows the real-space STS (f)
    and dI/dV (g) of a twisted graphene bilayer
    at 1.8$^{\circ}$,
    showing the emergence of van Hove
    singularities associated to
    non-abelian Landau levels.\cite{Li2009}}
    \label{fig:gauge_field}
\end{figure*}

\subsection{Generating flat bands from artificial gauge fields}
\label{sec:flatgauge}

A paradigmatic case of localized modes in a van der Waals material 
is non-uniform strained graphene (Fig. \ref{fig:gauge_field}a). The appearance of flat bands in this
system stems from the emergence of an artificial gauge field\cite{Vozmediano2010,PhysRevLett.121.146801,PhysRevLett.108.216802}.
The effect of strain is a create a term in the system Hamiltonian that mimics a magnetic field (``pseudo-magnetic field''). However, this differs from a real magnetic field as  the artificial
gauge field manifests as a positive magnetic field for electrons
in valley K and a negative electric field for electrons in valley K',
so that overall, the system does not break time-reversal symmetry.

The simplest instance of this is periodically rippled graphene
monolayers\cite{PhysRevLett.100.056807,Guinea2009,Guinea2009NF,RamezaniMasir2013,Amorim2016} (Fig. \ref{fig:gauge_field}b).
The emergence of the gauge field can be easily rationalized from the graphene
Hamiltonian\cite{Vozmediano2010}. For unstrained graphene, the low energy Hamiltonian
in a single valley takes the form\cite{RevModPhys.81.109}
$H = p_x \sigma_x + p_y \sigma_y $. In the presence of a global uniform
strain, the Dirac point get displaced from the K and K' points,
leading to Hamiltonian of the form
$H = (p_x+A_x) \sigma_x + (p_y+A_y) \sigma_y $.
Now, in a non-uniformly strained sample, we can
take that there is a local strain that changes
in real space, turning $A_x$ and $A_y$ spatially
dependent. Noting that $\vec A$ enters in the Dirac
Hamiltonian as a canonical momentum, we
can then identify a strain-induced
artificial magnetic field as
$\vec B = \nabla \times \vec A$.

Twisted graphene bilayers represent another case in which
spatial modulations give rise to an artificial gauge field\cite{PhysRevLett.108.216802,PhysRevB.99.155415}.
In twisted graphene bilayers, the stacking in space changes
between AA, AB and BA. The modulation of the interlayer
hopping due to the stacking brings up localized modes
stemming from gauge fields, that can be rationalized as follows.
A local Hamiltonian for a bilayer can be represented
by a $4\times4$ Hamiltonian, in which the off-diagonal
blocks contain the coupling between the two layers.
Due to the modulated stacking (Fig. \ref{fig:gauge_field}cd) the $2\times 2$ 
interlayer coupling is modulated in space. Given the
Dirac nature of the monolayer dispersion, that
modulated hopping can be rationalized as an
off-diagonal $2\times2$ gauge field, namely
a non-abelian $SU(2)$ gauge field\cite{PhysRevLett.108.216802}. 
This non-abelian
gauge field will thus give rise to associated pseudo
Landau levels, the lowest one of them being the
magic-angle flat bands at $1^\circ$\cite{PhysRevB.82.121407,Bistritzer2011,PhysRevB.99.155415}.

\subsection{Experiments flat bands from gauge fields}
The earliest experiments showing the emergence of pseudo-Landau levels
appeared in non-uniform strained graphene, which
naturally appear in graphene nanobubbles\cite{Levy2010}. In these setups
van Hove singularities in the density of states were shown to appear,
in contrast with the semimetallic spectra of unstrained samples.
The emergence of those resonances is associated to the emergent
gauge field, which was shown to correspond to up to an effective field of
300 T\cite{Levy2010}. These nanobubbles can also be realized with graphene deposited on a weakly interacting substrate, and in that case, the STM tip could be used to tune the strain and hence, the pseudo-magnetic field \cite{Mashoff2010,Georgi2017}. Finally, suspended graphene drumheads have also been used to investigate the effects of pseudomagnetic fields and how they can confine the charge carriers in graphene \cite{Klimov2012,PhysRevB.90.075426}.

The buckling of graphene monolayers can also be created by choosing
an appropriate substrate. In particular, recent experiments of graphene on top
of NbSe$_2$ showed that graphene gets a spontaneous buckling on this structure. Associated to the buckling, a periodic non-uniform strain appears in the graphene monolayer,
which gives rise to an elastic gauge field spontaneously\cite{Guinea2009,PhysRevB.91.161407,RamezaniMasir2013}. Signatures of pseudo-Landau
levels in this spontaneously buckled structure have been recently observed
with STM\cite{Mao2020}.

The second example of pseudo-Landau levels corresponds to twisted
bilayer graphene. Signatures of the lowest pseudo-Landau level,
usually known as magic angle flat bands
we observed early on, including
some signature of symmetry breaking\cite{Li2009,PhysRevLett.109.196802,PhysRevB.92.155409}. In particular,
these pseudo-Landau-levels show a strong localization
at the AA stacking regions of the twisted bilayer\cite{Li2009}.
Interestingly, higher index Landau levels 
can also lead to correlated states\cite{2019Lau}, show different
localization in the moire unit cell, and in particular
the next van Hove singularity shows a higher
extension around the AA regions\cite{Li2009,PhysRevLett.109.196802,PhysRevB.92.155409}.
Subsequent experiments have further
explored the nature of the lowest flat band,
in particular, observing spontaneous rotational
symmetry breaking due to electronic
interactions\cite{Jiang2019}.

\section{Outlook}
The possibility of artificial engineering states of matter with
van der Waals materials has demonstrated a huge potential
in the last few years. Beyond the instances of topological
insulators and superconductors, quantum spin-liquids and
flat band physics, their tunability opens prospects potentially
opening radical new directions in quantum matter.

Starting with topological superconductors, a challenge for
future experiments will be to braid the emergent Majorana modes,
in an analogous way as it has been proposed for semiconductor nanowires\cite{PhysRevX.6.031016,Alicea2011}.
The possibility of switching on and off topological
superconductivity with local gates
provides a direction for extending these
schemes to two-dimensional materials.
Furthermore, artificial engineering
can also allow engineering higher-order topological
superconductors, in which the braiding
of corner modes\cite{PhysRevResearch.2.032068,PhysRevResearch.2.043025}
can open up a potential new direction
for topological quantum computing with van der Waals materials.

Quantum spin-liquids open up exciting new experimental possibilities
well beyond their experimental confirmation. First, the emergence
of fractional spinon excitations in these systems opens possibilities
to controllable spinon transport, and ultimately
its interface with current spintronic devices\cite{RevModPhys.92.021003,Jungwirth2016}. Secondly, the emergence of
anyonic excitations in certain quantum spin
liquids\cite{PhysRevLett.125.227202,PhysRevX.10.031014} motivate potential future application of these systems
for topological quantum computing\cite{PhysRevX.10.031014}.

Flat band systems further offer novel possibilities
for emergent quantum matter, going beyond the well-known possibilities
for high-temperature superconductivity and symmetry broken states.
In particular, the emergence of topologically non-trivial flat bands
in twisted van der Waals materials provides an ideal starting
point for fractional quantum Hall states in the absence of magnetic
field, known as fractional Chern insulators\cite{PhysRevLett.106.236804,PhysRevLett.124.106803,PhysRevLett.126.026801,PhysRevResearch.2.023237,PhysRevResearch.2.023238}. Analogous phenomenology
for flat bands hosting spin-textured bands would further
provide playgrounds for fractional quantum spin Hall physics, a
state not found in nature yet. Ultimately, the combination of potential
fractional quantum Hall physics and superconductivity in twisted multilayers
provides an ideal starting point for engineering novel parafermion
states\cite{PhysRevX.4.011036,Alicea2016}. These states have resisted experimental realization so far
due to the difficulty of having simultaneously fractional quantum Hall
physics and superconductivity due to the large magnetic fields required.
Such limitation would, however, not exist for intrinsic fractional
quantum Hall states in graphene multilayers, providing an ideal solid
state platform for parafermion physics.

Advances in the last few years have drastically proved
the versatility of artificial engineering in van der Waals materials,
revealing a variety of exotic phenomena previously only observed in
rare compounds.
While many of those proposals require further materials engineering
and to further understand the physics of the underlying materials,
the steady development of the field suggest that some of those
goals may be achieved in the near future.

\begin{acknowledgments}
We thank our group members - past and present - and colleagues for inspiration and insightful discussions. We acknowledge support from the European Research Council (ERC-2017-AdG no.~788185 ``Artificial Designer Materials'') and Academy of Finland (Academy professor funding no.~318995 and Academy research fellow no.~331342).
\end{acknowledgments}

\bibliography{biblio}{}

\end{document}